\documentclass{ptapap}
\usepackage{amsmath}
\usepackage{amssymb}

\author{Gerald Handler}[CAMK]
\author{Rahul Jayaraman}[MIT]
\author{Donald W. Kurtz}[Mahi,UCL]
\author{Jim Fuller}[caltech]
\author{Saul A. Rappaport}[MIT]
\affil[CAMK]{Nicolaus Copernicus Astronomical Center, Polish Academy of Sciences, Bartycka 18, 00--716 Warsaw, Poland}
\affil[MIT]{MIT Department of Physics and MIT Kavli Institute for Astrophysics and Space Research, Cambridge, MA 02139, USA}
\affil[Mahi]{Centre for Space Research, Physics Department, North West University, Mahikeng 2745, South Africa}
\affil[UCL]{Jeremiah Horrocks Institute, University of Central Lancashire, Preston PR1 2HE, United Kingdom}
\affil[caltech]{TAPIR, California Institute of Technology, Pasadena, CA 91125, USA}
\title{Tidally Tilted Pulsators}

\begin{document}

\maketitle

\begin{abstract}

The tidally tilted pulsators are a new type of oscillating star in close binary systems that have their pulsation axis in the orbital plane because of the tidal distortion caused by their companion. We describe this group of stars on the basis of the first three representatives discovered and illustrate the basic methods used for their analysis. Their value for astrophysical study is rooted in the combination of the strengths of binary star and asteroseismic analyses; pulsational mode identifications can be achieved because the oscillations are visible over nearly 360 degrees of aspect throughout the orbital cycle. An illustrative case of a particularly interesting system is presented.

\end{abstract}

\section{Introduction}

\begin{figure}[h!]
\includegraphics[width=\textwidth,viewport=29 80 737 160]{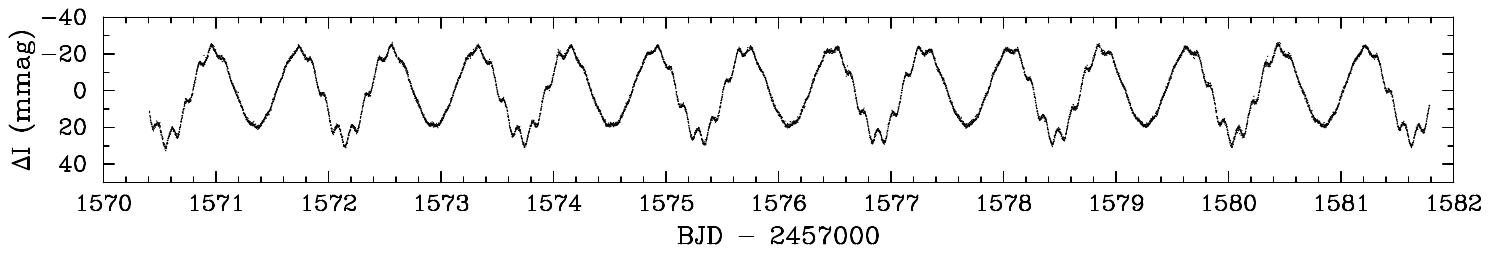}
\caption{Twelve-day segment of the {\it TESS} light curve of HD 74423.}
\label{fig:1}
\end{figure}

On 11 May 2019, American amateur astronomers Robert Gagliano and Thomas L. Jacobs reported an unusual light curve they noticed whilst visually surveying {\it TESS}\,\footnote{{\it TESS} is a NASA mission designed to discover hundreds of transiting planets smaller than Neptune with host stars bright enough for spectroscopic follow-up \citep{ricker}.} data (Fig.\,\ref{fig:1}).

This light curve, belonging to the 8$^{th}$ magnitude Southern Hemisphere star HD 74423, shows variability on two different time scales: a double-wave light variation with a period of about 1.6 d on which a 3-hr oscillation is superposed. Intriguingly, the amplitude of the shorter variation is modulated with the phase of the longer variation: it is largest near the deeper global minima, but is almost absent near the shallower global minima.

How can this be understood? The 1.6-d variation is typical of ellipsoidal variables, components of close binary systems distorted by the gravitational influence of a close companion \citep{morris}. Their light changes are predominantly due to the variable surface temperature of the observable hemisphere of the star. They therefore provide information on the orientation of the binary system towards the observer. The 3-hr oscillation, however, is due to stellar pulsation. HD 74423 is mid A star which puts it into the pulsational instability strip of $\delta$ Scuti stars \citep[e.g.][]{murphy}. It appears that the pulsations of HD 74423 are mostly confined to one of the stellar hemispheres.

How is that possible? There are three effects at play. Firstly, the tidal force exerted by the companion star pulls the pulsation axis of the oscillating star into the orbital plane. This is analogous to the rapidly oscillating Ap stars \citep{kurtz82} whose strong global magnetic fields tend to align the stellar pulsation axis with the magnetic axis instead of the rotation axis. Secondly, the oscillations are tidally trapped in certain parts of the star, in the case of HD 74423 the tidal poles. Thirdly, the pulsations are tidally amplified: the flux perturbations are stronger near the tidal poles where acoustic modes can propagate closer to the surface of the star, yet the two tidal poles can behave very differently. The theoretical framework describing tidally tilted pulsations has been developed by \cite{fuller}.

\section{The known tidally tilted pulsators}

Further study of HD 74423 \citep{handler20} revealed that this system consists of two almost identical chemically peculiar stars of the $\lambda$ Bootis type close to filling their Roche Lobes. The oscillating star has a single axisymmetric mode of pulsation that is directed towards its companion and has some ten times larger photometric amplitude in that direction.

A few months after the discovery of HD 74423, tidally tilted pulsation was discovered is the 5$^{th}$ magnitude star CO Cam. This is also an ellipsoidal system, but with a metallic-line A star primary with a somewhat shorter 1.3-d orbital period. The primary is not even close to filling its Roche Lobe, and the pulsations are multiperiodic, axisymmetric and also directed towards the companion \citep{kurtz20}.

\begin{figure}[h!]
\includegraphics[width=\textwidth,viewport=00 45 1563 438]{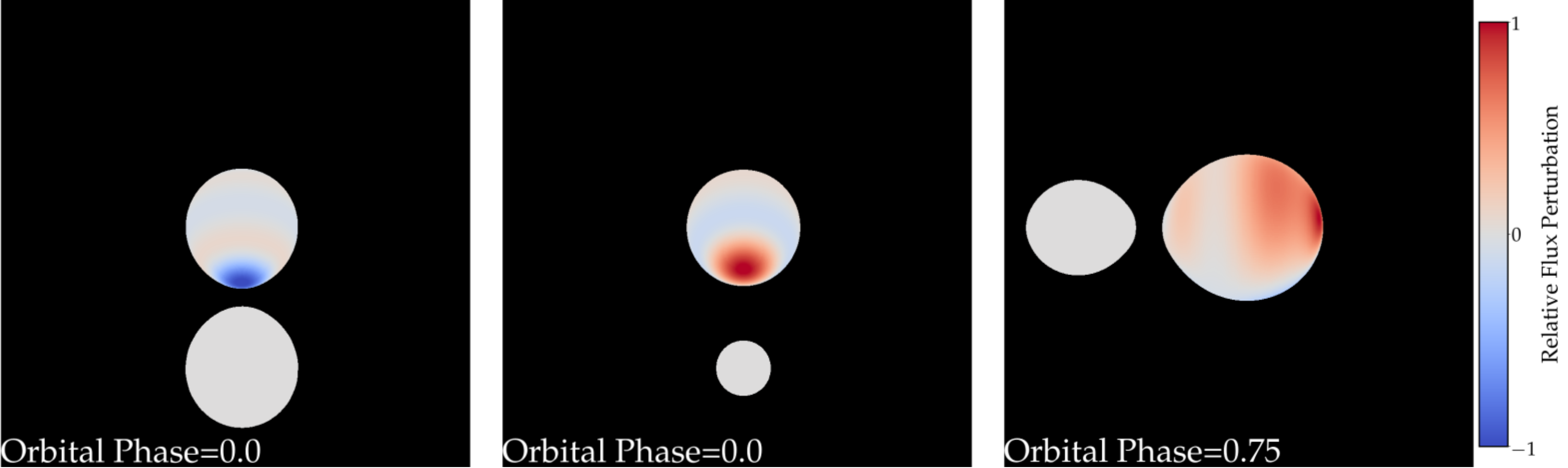}
\caption{The geometric configurations of three tidally tilted pulsators. Left: HD 74423, middle: CO Cam, right: TIC 63328020. Figure adapted from \cite{fuller}.}
\label{fig:2}
\end{figure}

The third star joining the ranks of tidally tilted pulsators is the 12$^{th}$ magnitude star TIC 63328020 \citep{rap21}. This object is again different from the other two, in that it has the shortest orbital period (1.1 d), has previously undergone mass transfer and appears chemically normal, is an eclipsing binary whose components are close to Roche-Lobe filling, but it oscillates mainly in a sectoral mode faced away from the companion star. Schematics of those systems are shown in Fig.\,\ref{fig:2}.

\subsection{Related objects in the literature}

The eclipsing binary V1031 Ori \citep{lee21} appears to have a short period tidally tilted oscillation. Further, the deep eclipsing binary U Gru \citep{bow19} shares some of the characteristics of the three stars discussed so far, and was therefore called a ``tidally perturbed pulsator''. However, it has some pulsation modes that do not appear to be affected by tides. Related cases are HL Dra \citep{shi21}, RR Lyn \citep{sou21} and RS Cha \citep{steindl}. Another eclipsing binary, TZ Dra, shows many tidally tilted oscillations that form regular patterns in the frequency spectra \citep{kahra21}. Whereas all these objects contain pulsators of the $\delta$ Scuti type, the stars with tidally perturbed pulsations in the systems V453 Cyg \citep{south20}, and VV Ori \citep{southetal} are of the $\beta$~Cephei type and hence considerably more massive.

\section{Analysing the tidally tilted pulsators}

\begin{figure}[h!]
\includegraphics[width=\textwidth,viewport=42 58 600 223]{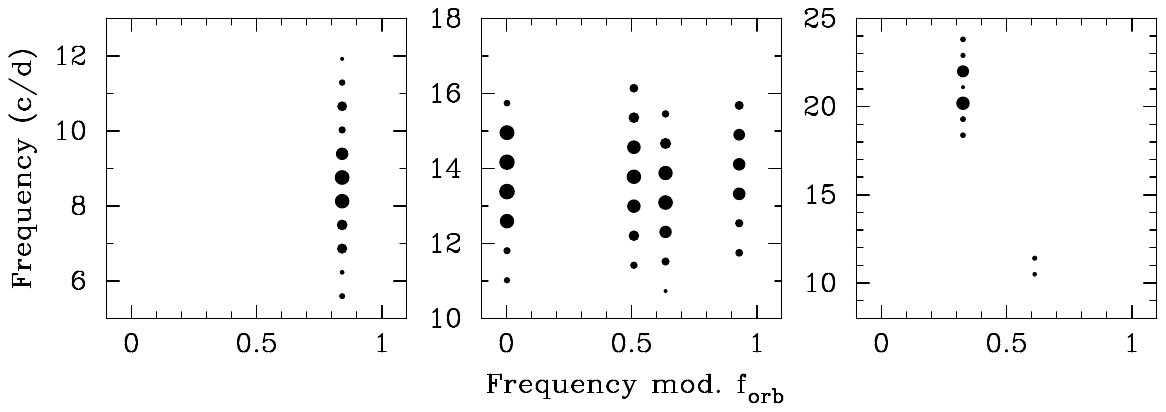}
\caption{Echelle Diagrams of the pulsation frequencies of HD 74423 (left), CO Cam (middle) and TIC 63328020 (right). The sizes of the dots are proportional to the oscillation amplitudes.}
\label{fig:3}
\end{figure}

An efficient tool to decide whether the pulsations of a star are possibly tidally tilted is an Echelle Diagram. Such a diagram plots the pulsation frequencies of a star relative to the frequency of some repetitive frequency spacing. As the pulsation amplitude and phase of tidally tilted pulsators are modulated with the orbital frequency, they are split into multiplets spaced with this orbital frequency in Fourier space. Consequently, frequencies corresponding to the same mode of oscillation in a tidally tilted pulsator will align in ridges in an Echelle Diagram relative to the orbital frequency. Figure \ref{fig:3} shows the Echelle Diagrams of the three objects discussed earlier.

\begin{figure}[h!]
\includegraphics[width=\textwidth,viewport=31 235 755 490]{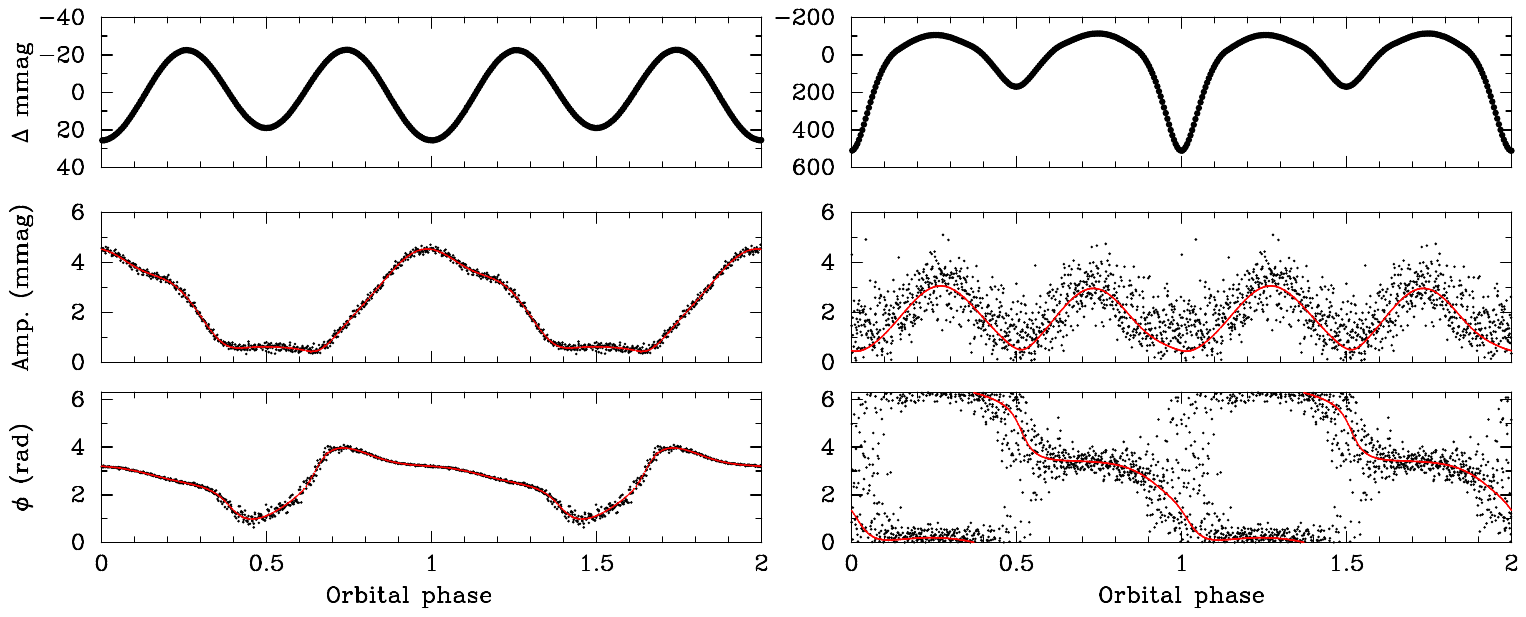}
\caption{The run of the pulsation amplitude and phases for HD 74423 (panels on the left-hand side) and TIC 63328020 (right-hand side) over the orbit. The top panels show the binary-induced light variations, the middle panels the pulsation amplitude and the lower panels the pulsation phase. Black dots are observational points, the red lines are fits derived from the Fourier decomposition of the light curves.}
\label{fig:4}
\end{figure}

Important also is the run of the pulsation amplitudes and phases over the orbit.  Figure \ref{fig:4} compares these quantities for HD 74423 and TIC 63328020. For HD 74423, the pulsation amplitude reaches its maximum value at the deeper ellipsoidal light minimum and its smallest value at the shallower minimum, as was already apparent from Fig.\ref{fig:1}. The situation is different for TIC 63328020, where maximum pulsation amplitude is observed near the ellipsoidal light maxima, and minimum pulsation amplitude plus a $\pi$ rad phase change at the times of eclipse, a result of the two hemispheres having opposite pulsation phases. The difference in this behaviour implies that different modes of pulsation are predominant in these two different stars. In fact, the surface shape of the underlying pulsation modes can thus be deduced as they are seen at variable aspect over the orbit.

This is not a new finding. Diagrams such as Fig.\ref{fig:4} have been used to identify the oscillation modes in rapidly oscillating Ap stars \citep[e.g.,][]{kur90}. In fact, \cite{reed05} have numerically predicted the Fourier spectra and light curves of pulsating stars with tilted pulsation axes for different modes of oscillation and different geometrical configurations of the binary systems. Whereas these methods can be applied in cases where the tidal distortion and tidal amplification of the pulsation modes are not too severe so that the angular pattern of the modes are well approximated by spherical harmonics. More general cases will need individual analysis along the lines described by \cite{fuller}. Nevertheless, it is important to realize that tidally tilted pulsators facilitate the identification of the observed modes of oscillation, a prerequisite for asteroseismic modeling.

\begin{figure}[h!]
\includegraphics[width=\textwidth,viewport=45 60 600 290]{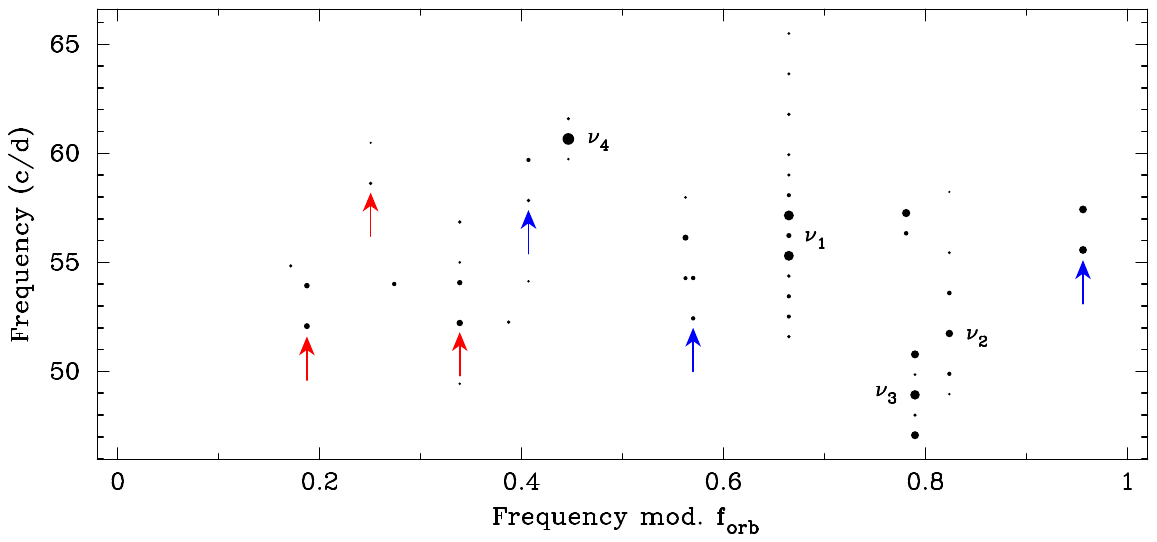}
\caption{Echelle Diagram of a recently discovered tidally tilted pulsator. The size of the dots is proportional to the signal amplitude. The centroids of four tidally tilted oscillation modes are marked ($\nu_1-\nu_4$). Furthermore six frequency doublets that are either in phase (red arrows) or are $\pi$ rad out of phase at primary minimum (blue arrows) are highlighted.}
\label{fig:5}
\end{figure}

\begin{figure}[ht]
\includegraphics[width=\textwidth,viewport=30 55 760 495]{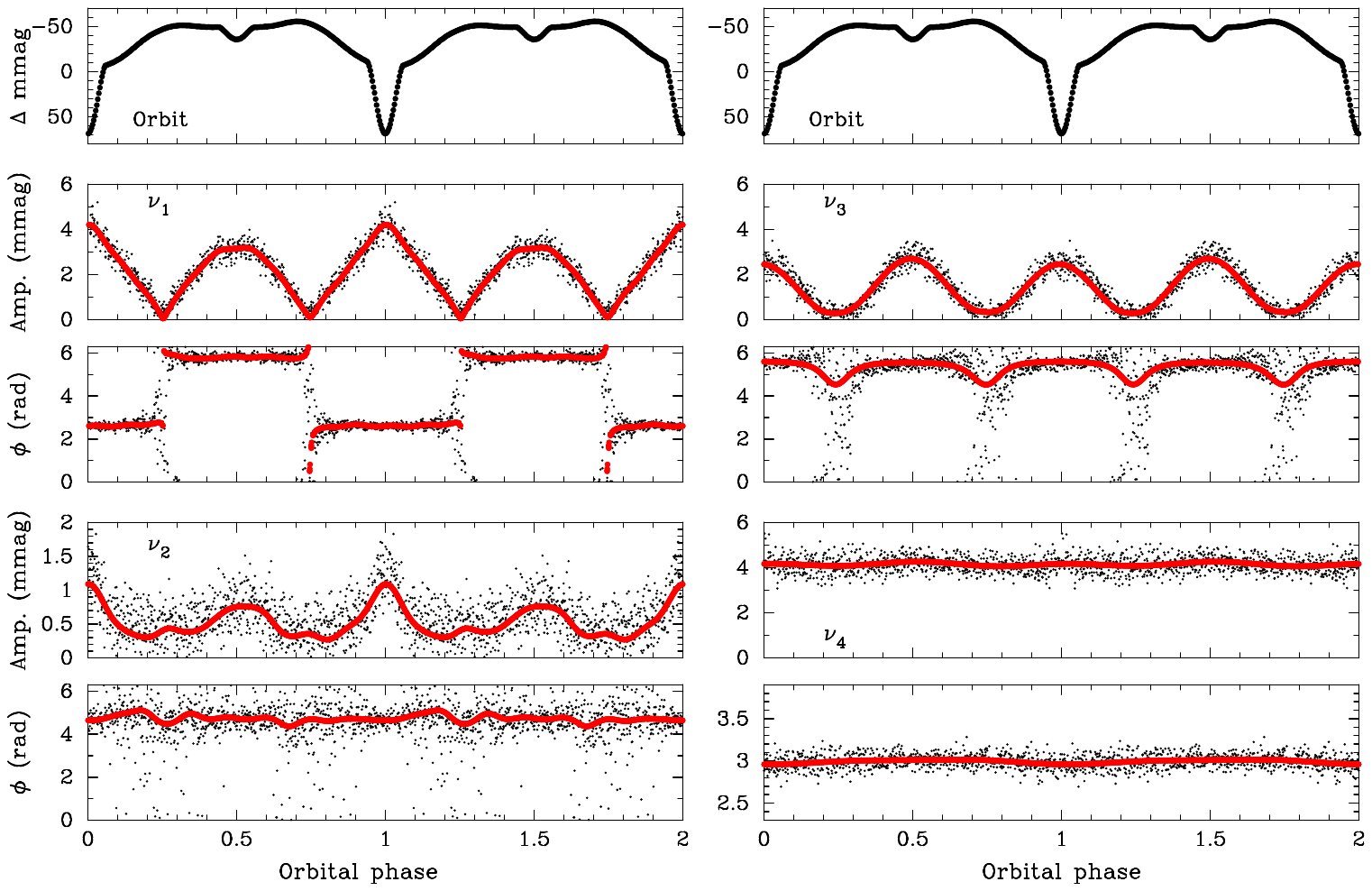}
\caption[]{The run of the pulsation amplitudes and phases of four selected modes of the star in Fig.\,\ref{fig:5} over the orbital cycle (bottom four panels left and right). The black dots are individual determinations, the red dots are running averages. Top: binary-induced light curve.}
\label{fig:6}
\end{figure}

Some tidally tilted pulsators may thus become available to detailed seismic analysis, provided that their pulsation spectra are rich enough. Such stars may already have been discovered. Figure\,\ref{fig:5} shows an Echelle Diagram of the pulsation frequencies of a tidally tilted pulsator identified more recently\footnote{This is work in progress, hence this object shall remain nameless for the time being.}.
The structures within the different vertical ridges are markedly different which points towards pulsation modes with different spherical degree $l$ and azimuthal order $m$. Figure\,\ref{fig:6} shows the run of pulsation amplitude and phase of four strong modes over the orbit.

The behaviour of these modes is varied. Mode $\nu_4$ shows no amplitude modulation (except near the eclipses) and almost no phase modulation. It is therefore likely radial. Mode $\nu_1$ has maximum amplitude at the eclipse minima and $\pi$ rad phase jumps at quadrature: it is mostly an axisymmetric dipole. Modes $\nu_2$ and $\nu_3$ require more in-depth analysis. Of interest are also the frequency doublets pointed out in Fig.\,\ref{fig:5}. Those with high/low amplitudes at primary minimum are good candidates for axisymmetric modes, whereas those with high/low amplitudes at the light maxima would most likely be sectoral modes \citep{reed05}. For this star, a complete mode identification appears to be within reach. Observations are in progress to determine its spectroscopic orbit. In combination with binary modelling it should be possible to arrive at a complete description of the system, including the geometries of the binary orbit and the pulsations, both constraining the orbital inclination.

\section{Concluding remarks}

This article described the first representatives of a new type of pulsating stars in binary systems, whose pulsations are affected by the tidal distortion of their close companions. These are called tidally tilted pulsators and their main feature is that they have their pulsation axis pulled into the orbital plane. More special cases are casually called single sided pulsators that additionally have the pulsation modes distorted and the pulsation amplitudes amplified in certain parts of the star.

Tidally tilted pulsators offer new insights into stellar astrophysics as they are oblique pulsators: their oscillation modes are visible under variable aspect over the orbit which provides clues towards mode identification. The difficulty of deriving mode identifications is a nagging problem in asteroseismology of pulsating main sequence stars driven by the $\kappa$ mechanism, which can thus be overcome for tidally tilted pulsators. In addition, accurate stellar parameters from binarity provide tight constraints to be used in the modelling process.

The tidally tilted pulsators known so far share little except their very nature: they have different amounts of tidal distortion, different chemical compositions, different pulsational behaviour, and different evolutionary histories. Discoveries of more representatives of these stars are needed to enable their systematic study.



\acknowledgements{We are indebted to the Visual Search Group (Andrew Vanderburg, Thomas L. Jacobs, Robert Gagliano, Martti H. Kristiansen, Mark Omohundro, Hans Martin Schwengeler, Daryll M. LaCourse and Ivan A. Terentev) for pointing us to many interesting objects, making the discovery of tidally tilted pulsators possible. We thank David Mkrtichian for pointing out a mistake in the original manuscript that is now fixed. GH acknowledges financial support by the Polish NCN grant 2015/18/A/ST9/00578.}

\bibliographystyle{ptapap}
\bibliography{ghandler}

\end{document}